\documentclass[12pt]{article}
\usepackage{amssymb}

\usepackage{amsmath}

\input{tcilatex}

\newcommand{\benumerate}{\begin{enumerate}}
\newcommand{\eenumerate}{\end{enumerate}}

\newcommand{\bitemize}{\begin{itemize}}

\newcommand{\eitemize}{\end{itemize}}
\newcommand{\der}[2]{\frac{\partial #1}{\partial #2}}

\newcommand{\paperJPA}[5]{#1 #2 {\em #3} {\bf#4} #5}

\newcommand{\bookJPA}[4]{#1 #2 {\em ``#3''} (#4)}

\begin{document}

\date{}

\title{Geometrical optics in nonlinear media and integrable equations}

\author{Boris G. Konopelchenko and Antonio Moro  \\
\small{Dipartimento di Fisica dell'Universit\`{a} di Lecce
and INFN,} \\ \small{Sezione di Lecce, via Arnesano, I-73100 Lecce, Italy} \\
\small{E-mail: konopel@le.infn.it and antonio.moro@le.infn.it}}

\maketitle

\begin{abstract}
It is shown that the geometrical optics limit of the Maxwell equations
for certain nonlinear media with slow variation along one axis and
particular dependence of dielectric constant on the frequency and
fields gives rise 
to the dispersionless Veselov-Novikov equation for refractive
index. It is demonstrated 
that the last one is amenable to the quasiclassical
$\bar{\partial}$-dressing method. A connection is noted between
geometrical optics phenomena under consideration and quasiconformal
mappings on the plane.

\end{abstract}

Great variety of nonlinear phenomena from various fields of physics,
applied physics, mathematics and applied mathematics can be modeled
by nonlinear integrable equations \cite{Zakharov}-\cite{Shiota}. A subclass of such
equations , the so-called dispersionless integrable equations, has
attracted a particular interest during the last
decade \cite{Zakharov2}-\cite{Wiegmann}.

In this letter, we will show that the propagation of electromagnetic
waves of high frequency in certain nonlinear media is governed by the
dispersionless Veselov-Novikov (dVN) equation. Namely, we will
demonstrate that the Maxwell equations describing the propagation of
waves in media characterized by slow variation along the axis $z$ and
particular dependence of dielectric constant on frequency and fields in the limit
of geometrical optics give rise to the dVN equations for refractive
index.
These equations provide us with integrable deformations of the plane
eikonal equation which preserve, in particular, the total ``plane''
squared refractive index $\int\int n^{2} dx dy$.
Under more special conditions one gets the dispersionless
Kadomtsev-Petviashvili (dKP) equation. We will show that the dNV
equations are treatable by the quasiclassical $\bar{\partial}$-dressing
method developed recently in \cite{Konopelchenko2}-\cite{Konopelchenko4}. The quasiclassical
$\bar{\partial}$-dressing method reveals also a connection between
the geometrical optics phenomena under consideration and the theory of
quasiconformal mappings on the plane.

We start with the Maxwell equations in the absence of sources (we
put the velocity of the light $c=1$)
\begin{eqnarray}
\label{maxwell}
&&\nabla \wedge {\bf H} - {\bf \dot{D}}=0~~~~~~\nabla \cdot {\bf D} = 0 \nonumber \\
&&\nabla \wedge {\bf E} + {\bf \dot{B}}=0 ~~~~~~~\nabla \cdot {\bf B} = 0
\end{eqnarray}
and the material equations of a medium
\begin{equation}
\label{materials}
{\bf D} = \epsilon {\bf E},~~~~~~{\bf B} = \mu {\bf H}.
\end{equation}
Here and below $\nabla = \left(\partial_{x},\partial_{y},\partial_{z}
\right)$, while $\cdot$ and $\wedge$ denote the scalar and vector
product respectively.   

We will study a propagation of electromagnetic waves of the fixed,
high frequency $\omega$, i.e. we will look for the solutions of the
Maxwell equations of the form \cite{Born}
\begin{eqnarray}
\label{full_fields}
&&{\bf E}\left(x,y,z,t\right) = {\bf E_{0}}\left(x,y,z\right)
e^{-i\omega t + i \omega S\left(x,y,z\right)} \nonumber\\
&&{\bf H}\left(x,y,z,t\right) = {\bf H_{0}}\left(x,y,z\right) 
e^{-i\omega t + i \omega S\left(x,y,z\right)},
\end{eqnarray} 
where ${\bf E_{0}}$, ${\bf H_{0}}$ and the phase $S\left(x,y,z
\right)$ are certain functions.\\ 
In addition we assume that the nonlinear medium is characterized 1)
by the independence of the magnetic permeability  $\mu(x,y,z)$ on the
frequency $\omega$; 2) by the Cole-Cole dependence \cite{ColeCole}
\begin{equation}
\epsilon = \epsilon_{0} + \frac{\tilde{\epsilon}}{1+\left (i
  \omega \tau_{0} \right)^{2\nu}},~~~~~0<\nu <\frac{1}{2}
\end{equation}
of the dielectric constant $\epsilon$ on $\omega$ (where
$\epsilon_{0}(x,y,z)$ and $\tau_{0}$ are independent on $\omega$ and
$\tilde{\epsilon}$ is a function of coordinates and fields); 3) by
slow variation of all quantities along the axis $z$ such
that $\der{}{z} = \omega^{-\nu} \der{}{\xi}$, where $\xi$ is a ``slow''
variable defined by $z = \omega^{\nu} \xi$, i.e.
\begin{eqnarray}
\label{dependence_fields}
S &=& S\left(x,y,\xi \right) \nonumber \\
{\bf E_{0}} &=& {\bf E_{0}}\left(x,y,\omega^{\nu} \xi \right) \nonumber \\
{\bf H_{0}} &=& {\bf H_{0}}\left(x,y,\omega^{\nu} \xi \right).
\end{eqnarray}  
Now, rewriting equations~(\ref{maxwell}) and~(\ref{materials}) as the
second order differential equations for the electric and magnetic
fields, respectively, using (\ref{full_fields}) and
(\ref{dependence_fields}) and taking into account that at 
$\omega \rightarrow \infty$ 
\begin{equation}
\epsilon = \epsilon_{0} + \frac{\epsilon_{1}}{\omega^{2 \nu}}
\end{equation}
where $\epsilon_{1} = \left(i \tau_{0}\right)^{2 \nu}
\tilde{\epsilon}$, one gets in the leading $\omega^{2}$ order 
\begin{equation}
\label{eikonal}
S_{x}^{2} + S_{y}^{2} = \mu \epsilon_{0}
\end{equation}
while, in the next $\omega^{2-2\nu}$ order, one obtains
\begin{equation}
\label{z_direction}
S_{\xi} = \left(\mu \epsilon_{1}\right)^{\frac{1}{2}}.
\end{equation}
 
Equation~(\ref{eikonal}) is the standard eikonal equation on the $x,y$
plane with the refractive index $n\left(x,y,\xi \right) = \left(\mu
\epsilon_{0} \right)^{\frac{1}{2}}$.
Slow variation of $S$ along $z$-axis is defined by
equation (\ref{z_direction}), where $\mu \epsilon_{1}$ depends both on 
coordinates $x,y,\xi$ and fields. We also assume that the
absorption effects are negligible, so $\mu \epsilon$ can be taken as
real one . We would like to note that the
first two features of the medium mentioned above are quite generic
and they are valid for a variety of dielectric media \cite{ColeCole}-\cite{ColeRecent2}, while the particular type of slow variations along $z$-axis
given by~(\ref{z_direction}), seems, did not attract attention
before.

Since the function $\epsilon_{1}$ is determined by the
geometrical optics limit $\omega \rightarrow \infty$ and the time
translation symmetry $t \rightarrow t + const $ of the Maxwell
equations for the solutions of the type~(\ref{full_fields}) is
equivalent to the symmetry under transformations $S \rightarrow S +
const$, one concludes that the function $\epsilon_{1}$ for nonlinear
media may depend on the coordinates $x,y,\xi$ and only on $S_{x}$ and
$S_{y}$.
Thus, the geometrical optics limit of the Maxwell equations in nonlinear
media under consideration is governed by the equations
\begin{eqnarray}
\label{pair1_1}
&&S_{x}^{2} + S_{y}^{2} = n^{2}\left(x,y,\xi \right),  \\
\label{pair1_2}
&&S_{\xi} = \varphi \left(x,y,\xi;S_{x},S_{y}\right)
\end{eqnarray}
where $\varphi$ is some real-valued function.
Introducing the variables $z = x+iy$, $\bar{z} = x-iy$ (please, do not confuse
it with the variable used in equation~(\ref{full_fields})), one
rewrites these equations as follows 
\begin{eqnarray}
\label{eikonal_complex}
&& S_{z} S_{\bar{z}} = u(z,\bar{z},\xi) \\
\label{z_direction_complex}
&& S_{\xi} = \varphi\left (z,\bar{z};\xi,S_{z},S_{\bar{z}} \right),
\end{eqnarray} 
where $u = 4 n^{2}$. \\
The compatibility of equations~(\ref{pair1_1}) and~(\ref{pair1_2}) (or
~(\ref{eikonal_complex}), (\ref{z_direction_complex})) imposes
constraints on the possible forms of the function $\varphi$, namely
\begin{equation}
\label{compatibility}
S_{\bar{z}} \der{\varphi}{z} + S_{z}
\der{\varphi}{\bar{z}} + u_{z} \varphi' + u_{\bar{z}}
\varphi'' = u_{\xi},
\end{equation}
where 
\begin{equation}
\varphi' =
\der{\varphi}{S_{z}}\left(z,\bar{z};S_{z},S_{\bar{z}}
\right),~~~~~~\varphi'' =
\der{\varphi}{S_{\bar{z}}}\left(z,\bar{z};S_{z},S_{\bar{z}}
\right).
\end{equation}
Here we restrict ourself by functions $\varphi$ which are polynomial
in $S_{z}$, $S_{\bar{z}}$ and compatible with real-valuedness of $S$ and $u$.
For the simplest choice $\varphi = \alpha_{0}(z,\bar{z},\xi)$,
equation~(\ref{compatibility}) obviously gives $\alpha_{0}=const$,
i.e. $u_{\xi}=0, S = \alpha_{0} \xi +
\tilde{S}(z,\bar{z})$. For the linear function $\varphi =
\alpha S_{z} + \bar{\alpha} S_{\bar{z}}+\beta+ \bar{\beta}$, one gets
$\alpha = \alpha(z), \beta=const$, and 
\begin{equation}
\label{linear}
u_{\xi} = \left(\alpha u \right)_{z} + \left(\bar{\alpha} u
\right)_{\bar{z}}.
\end{equation}
 For the quadratic $\varphi = \alpha S_{z}^{2} + \bar{\alpha}
S_{\bar{z}}^{2}+\beta S_{z}+ \bar{\beta} S_{\bar{z}}+
\gamma + \bar{\gamma}$, equations~(\ref{compatibility})
and~(\ref{eikonal_complex}) imply $\alpha = 0$, $\beta = \beta(z)$ and
$\gamma = const$, i.e. one gets the previous linear case.

The cubic $\varphi = \alpha S_{z}^{3} + \bar{\alpha} S_{\bar{z}}^{3}+ \beta S_{z}^{2}+\bar{\beta} S_{\bar{z}}^{2}+
\gamma S_{z} + \bar{\gamma} S_{\bar{z}} + \delta + \bar{\delta}$
obeys equations~(\ref{compatibility}) and~(\ref{eikonal_complex}) if
\begin{eqnarray}
&&\alpha = \alpha(z),~~~~~~\beta = 0, \nonumber \\
&&\gamma_{\bar{z}} = - \alpha_{z} u - 3 \alpha u_{z},
\nonumber \\
&&\delta = const,
\end{eqnarray}
and one gets
\begin{equation}
\label{cubic}
u_{\xi} = \left(\gamma u\right)_{z} + \left(\bar{\gamma} u
\right)_{\bar{z}}. 
\end{equation}
In the particular case $\alpha=1$ and,
consequently $\gamma_{\bar{z}} = -3 u_{z}$, equation~(\ref{cubic}) is nothing
but the dispersionless Veselov-Novikov (dVN) equation introduced
in \cite{Krichever,Konopelchenko4}.

In a similar manner one can construct nonlinear equations which
correspond to higher degree polynomials $\varphi$. These higher degree
cases apparently become physically relevant for the phenomena with
large values of $S_{x}$ and $S_{y}$.
Thus, if we formally admit all possible degrees of $S_{z}$ and
$S_{\bar{z}}$ in the right hand side of
equation~(\ref{z_direction_complex}), then one has an infinite family
of nonlinear equations, which may govern the $\xi$-variations of the wave
fronts and ``refractive index'' $u$. Since
equation~(\ref{z_direction_complex}) should respects the
symmetry $S \rightarrow - S$ of the eikonal
equation~(\ref{eikonal_complex}), one readily concludes that only polynomials $\varphi$ of
the form $\sum_{n=1}^{N} u_{n} S_{z}^{2n-1} + \sum_{n=1}^{N}
\bar{u}_{n} S_{\bar{z}}^{2n-1}$ , are
admissible (the constant terms which have appeared in the
cases~(\ref{linear}), (\ref{cubic}) discussed above are, in fact,
irrelevant). In the case $u_{N}=1$ one gets the dVN equation mentioned above ($N=2$) and
the so-called dVN hierarchy of nonlinear equations. The dVN equation
has been introduced in \cite{Krichever,Konopelchenko4} as the dispersionless limit of the VN
equation, which is the $2+1$-dimensional integrable
generalization of the famous Korteweg-de-Vries (KdV)
equation \cite{Zakharov}-\cite{Jimbo},\cite{Konopelchenko}.

Let us consider a more specific situation in which the
propagation of electromagnetic waves in the media discussed above
exhibits also a slow variation along the axis $y$, namely $\partial_{y}
= \epsilon \partial_{\eta}$, where $\eta$ is a slow variable defined
by $y = \frac{\eta}{\epsilon}$ and $\epsilon$ is a small
parameter. If one assumes that the functions $S$ and $u$ in the
eikonal equation~(\ref{pair1_1}) have the following behavior at small
$\epsilon$
\begin{eqnarray}
\label{y_direction}
&&S = \frac{y}{2 \epsilon} - \tilde{S}\left(x,\eta,\xi \right)\nonumber \\
&&u\left(x, \frac{\eta}{\epsilon},\xi\right) = \frac{1}{4 \epsilon^{2}} -
q\left(x,\eta,\xi \right)
\end{eqnarray}
and the polynomial
$\varphi\left(x,\frac{\eta}{\epsilon},\xi;S_{x}, \epsilon S_{\eta}
\right)$ has an appropriate behavior as $\epsilon \rightarrow 0$, then
in this limit, equations~(\ref{pair1_1}) and~(\ref{pair1_2}) are reduced to
\begin{eqnarray}
\label{eikonal_KP}
&&\tilde{S}_{\eta} = \tilde{S}_{x}^{2} + q \nonumber\\
&&\tilde{S}_{\xi} = \tilde{\varphi}\left(x,\eta,\xi;\tilde{S}_{x} \right)
\end{eqnarray} 
where $\tilde{\varphi}$ is an odd order polynomial in $\tilde{S}_{x}$.
Equations~(\ref{eikonal_KP}) describe propagation of the quasi-plane
wave fronts $y=const$ in a medium with very large refractive index.
Compatibility of equations~(\ref{eikonal_KP}) gives rise to the well
known dispersionless Kadomtsev-Petviashvili (dKP) equation $q_{\xi} =
\frac{3}{2}q q_{x} + \frac{3}{4} \partial_{x}^{-1} q_{\eta \eta}$ and the whole dKP
hierarchy. The dKP equation is rather well studied (see e.g. \cite{Krichever,Singular} and
reference therein). The KP equation itself represents the most known
2+1-dimensional integrable generalization of the KdV equation.

The dVN and the dKP equations being relevant in particular situations
of propagation of waves in certain nonlinear media have an advantage
to be solvable. Various methods have been applied to solve the dKP
equation, including the quasiclassical $\bar{\partial}$-dressing
method \cite{Konopelchenko2}-\cite{Konopelchenko4}.
Here we will demonstrate that the dVN equation is amenable to this
method too. \\
The quasiclassical $\bar{\partial}$-dressing method is
based on the nonlinear Beltrami equation \cite{Konopelchenko2}-\cite{Konopelchenko4}
\begin{equation}
\label{Beltrami_nonlinear}
S_{\bar{\lambda}} = W\left(\lambda,\bar{\lambda};S_{\lambda} \right),
\end{equation}
where $\lambda$ is the complex variable, $S_{\lambda} =
\der{S}{\lambda}$ and $W$ is a certain function ($\bar{\partial}$-data).
To construct integrable equations one has to specify the domain $G$ (in
the complex plane $\mathbb{C}$) of
support for the function $W$ ($W=0$, $\lambda \in \mathbb{C}\setminus G$) and look for solution of~(\ref{Beltrami_nonlinear})
in the form $S = S_{0} + \tilde{S}$, where the function $S_{0}$ is
analytic inside $G$, while $\tilde{S}$ is analytic outside $G$ \cite{Konopelchenko2}-\cite{Konopelchenko4}.
To get the dVN equation and the whole dVN hierarchy, we choose $G$ as
the ring ${\cal D} = \left \{\lambda \in \mathbb{C} :~ \frac{1}{a} <
|\lambda| < a \right\}$, where $a$ is an arbitrary real number ($a >
1$), and the function $S_{0}$ in the form 
\begin{equation}
\label{asymp_S0}
S_{0} = \sum_{n=1}^{\infty}x_{n} \lambda^{2n-1} +
\sum_{n=1}^{\infty}\bar{x}_{n} \lambda^{-2n+1}.
\end{equation}
We also assume that
\begin{equation}
\label{asymp_Stilde_infinity}
\tilde{S} = \sum_{n=0}^{\infty}
\frac{S_{2n+1}^{(\infty)}}{\lambda^{2n+1}};~~~~~~\lambda \rightarrow \infty
\end{equation}
``normalization'' and
\begin{equation}
\label{asymp_Stilde_zero}
\tilde{S} = \sum_{n=0}^{\infty} \lambda^{2n+1}
S_{2n+1}^{(0)};~~~~~~\lambda \rightarrow 0.
\end{equation}
Note that a function $S\left(\lambda,\bar{\lambda},x_{n},\bar{x}_{n}\right)$
has the properties~(\ref{asymp_S0}-\ref{asymp_Stilde_zero}) if it
obeys the constraints
\begin{eqnarray}
\label{parity}
&&S\left(-\lambda,-\bar{\lambda} \right) = -S\left(\lambda,\bar{\lambda}
\right), \\ 
&&\label{parity_circle}
\bar{S}\left(\lambda,\bar{\lambda}\right) = S\left(\frac{1}{\bar{\lambda}},\frac{1}{\lambda} \right).
\end{eqnarray}
The constraint~(\ref{parity_circle}) also implies that
\begin{equation*}
S_{2n+1}^{(\infty)} = \bar{S}_{2n+1}^{(0)}.
\end{equation*}
An important property of the nonlinear
$\bar{\partial}$-problem~(\ref{Beltrami_nonlinear}) is that the
derivatives of $S$ with respect to any independent variable $x_{n}$
obeys the linear Beltrami equation
\begin{equation}
\label{Beltrami}
\left(\der{S}{x_{n}} \right)_{\bar{\lambda}}
=W'\left(\lambda,\bar{\lambda};S_{\lambda}\right) \left(\der{S}{x_{n}}
\right)_{\lambda}, 
\end{equation} 
where $W'\left (\lambda,\bar{\lambda};\phi \right) =
\der{W}{\phi}\left(\lambda,\bar{\lambda};\phi
\right)$. Equations~(\ref{Beltrami}) has two basic properties,
namely, 1) any differentiable function of a solution is again a
solution; 2) under certain mild conditions on $W'$, a solution which is
equal to zero at certain point $\lambda_{0} \in \mathbb{C}$, vanishes identically
(Vekua's theorem) \cite{Vekua}.
The use of these two properties allows us to construct algebraic
equations of the form $\Omega\left(x_{n},\bar{x_{n}},S_{x_{n}},S_{\bar{x}_{n}}
\right) = 0$. 

Denoting $x_{1} = z$, choosing $x_{2} = \bar{x}_{2} = \xi$, taking into
account that $S_{z} = \lambda + \tilde{S}_{z}$, $S_{\bar{z}} =
\lambda^{-1} + \tilde{S}_{\bar{z}}$, $S_{\xi} =
\left(\lambda^{3}+\frac{1}{\lambda^{3}} \right ) + \tilde{S}_{\xi}$,
and using the properties of the linear Beltrami
equation~(\ref{Beltrami}) mentioned above, one gets 
\begin{eqnarray}
\label{eikonal2}
&&S_{z} S_{\bar{z}} = u, \\
\label{z_direction2}
&&S_{\xi} = S_{z}^{3} + S_{\bar{z}}^{3} + V S_{z} + \bar{V} S_{\bar{z}},
\end{eqnarray}
where
\begin{equation}
u\left(z,\bar{z}, \xi \right) = 1 + S_{1\bar{z}}^{(\infty)}, ~~~~~~V =
-3 S_{1 z}^{(\infty)} = -3 \partial_{\bar{z}}^{-1}u_{z}.
\end{equation}
Evaluating the terms of the order $\frac{1}{\lambda}$ in the both
sides of equation~(\ref{z_direction2}), one gets the dVN equation
\begin{equation}
\label{dVN}
u_{\xi} = - 3 \left(u \partial_{\bar{z}}^{-1} u_{z} \right)_{z} - 3
\left(u \partial_{z}^{-1}u_{\bar{z}} \right)_{\bar{z}}.
\end{equation}
Considering higher ``times'' $x_{3},x_{4},\dots$ such that $x_{n} =
\bar{x}_{n}$, one obtains the equations
\begin{equation}
\label{higher}
S_{x_{n}} = \sum_{m=1}^{n} \left(u_{m} S_{z}^{2m-1} + \bar{u}_{m} S_{\bar{z}}^{2m-1}\right), ~~~~~n=3,4,\dots
\end{equation}
Equations~(\ref{higher}) together with~(\ref{eikonal2}) and~(\ref{z_direction2}) provides us
with the infinite dVN hierarchy of equations. 

Thus, the quasiclassical
$\bar{\partial}$-dressing method allows us to treat the eikonal
equation~(\ref{eikonal2}), equations (\ref{z_direction2})
and~(\ref{higher}) which describe its deformations and the
corresponding dVN hierarchy for $u$, in a way similar to the dKP and
d2DTL hierarchies \cite{Konopelchenko2}-\cite{Konopelchenko4}. 
These integrable dVN deformations of the eikonal
equation~(\ref{eikonal2}) are quite special. In particular, they are
characterized by the existence of an infinite set of integral
quantities (integrals of motion for dVN equation), which are
preserved by these deformations. The simplest of them, for $u$ such
that $u \rightarrow 0$ as $|z| \rightarrow \infty$, is given by the
total squared refractive index $C_{1}
= \int \int u\left(z,\bar{z},\xi \right) dz \wedge d\bar{z}$
($C_{1\xi} = 0$). \\

Constraint~(\ref{parity_circle}) guarantees
that the refractive index $u$ is a real one. Indeed, taking the
complex conjugation of equation~(\ref{eikonal2}), using the
differential consequences (with respect to $z$ and $\bar{z}$), of the
above constraint and taking into account the independence of the
l.h.s. of equation~(\ref{eikonal2}) on $\lambda$,$\bar{\lambda}$, one
gets
\begin{equation}
\bar{u}\left (x_{n}\right) =
\overline{S_{z}\left(\lambda,\bar{\lambda},x_{n} \right)}~~
\overline{S_{\bar{z}}\left(\lambda,\bar{\lambda},x_{n} \right)} =
S_{\bar{z}}\left(\frac{1}{\bar{\lambda}},\frac{1}{\lambda},x_{n} \right)
S_{z}\left(\frac{1}{\bar{\lambda}},\frac{1}{\lambda},x_{n} \right) = u\left(x_{n} \right). 
\end{equation}
Moreover, this constraint implies that the function $S$ is real-valued
on the unit circle $\left|\lambda \right| = 1$ ($\bar{S}
\left(\lambda,\bar{\lambda} \right) =
S\left(\lambda,\bar{\lambda}\right)$, $|\lambda| =1$). \\
Thus, the approach proposed provides us with the
real-valued refractive index  $u\left(x_{n}\right)$ and the phase
function $S$ together with their slow variations in the direction
orthogonal to the plane $x,y$.
Note that the refractive index $u = 4 n^{2}$ not necessary should be
positive. For certain media the product $\mu \epsilon_{0}$ can be
negative as well \cite{Veselago}. \\
The quasiclassical $\bar{\partial}$-dressing method provides also an
effective method to construct exact solutions of the dVN equations as
in the dKP case \cite{Konopelchenko2}-\cite{Konopelchenko4}. Exact
solutions of the dVN equation and their applications in geometrical
optics will be discussed in a separate paper.

Finally we would like to note that solutions of the nonlinear Beltrami
equation~(\ref{Beltrami_nonlinear}) represent themself the so-called
quasiconformal mappings (for dKP equation see \cite{Konopelchenko2}). In
the our case 
we have quasiconformal mappings of the ring ${\cal D}$, given by the function $S\left(\lambda,\bar{\lambda},x_{n}
\right)$, with the boundary
conditions~(\ref{asymp_Stilde_infinity}) and~(\ref{asymp_Stilde_zero})
and constraints~(\ref{parity}) and~(\ref{parity_circle}).
Thus the quasiclassical $\bar{\partial}$-dressing approach to the
eikonal equation~(\ref{eikonal2}) and its deformations by dVN
hierarchy establishes a connection between the geometrical optics and
the theory of the quasiconformal mappings on the plane
\cite{Ahlfors,Letho}. 
\bigskip
{\bf Acknowledgment}\\
This work is supported in part by the COFIN PRIN `Sintesi' 2002.

\end{document}